\documentclass{aa}
\usepackage[pdftex]{graphicx}
\usepackage{txfonts}
\usepackage{longtable}
\usepackage{pdflscape}
\usepackage{natbib}
\def \etal {\textit{et al. }}
\def \RpIn {$13.49 $}
\def \RpInErr {$0.55 $}
\def \RpOut {$6.14 $}
\def \RpOutErr {$0.29 $}
\def \MpIn {$324.2 $}
\def \MpInErr {$8.8 $}
\def \MpOut {$6.4 $}
\def \MpOutErr {$0.8 $}
\def \TeqIn {$478.1 $}
\def \TeqOut {$403.3 $}
\def \RpIII {$1.68 $}
\def \RpIIIErr {$0.17 $}
\def \RpIV {$1.93  $}
\def \RpIVErr {$0.19 $}
\def \RhoIn {$0.729 $}
\def \RhoInErr {$0.026 $}
\def \RhoOut {$0.152 $}
\def \RhoOutErr {$0.019 $}

\begin{document}
    \title{An Independent Planet Search In The {\emph Kepler} Dataset}
 	\subtitle{II. An extremely low-density super-Earth mass planet around Kepler-87.}
    \author{Aviv Ofir \inst{1,2} \and Stefan Dreizler \inst{1} \and Mathias Zechmeister \inst{1} \and Tim-Oliver Husser \inst{1}}

    \institute{Institut f\"ur Astrophysik, Georg-August-Universit\"at, Friedrich-Hund-Platz 1, 37077 G\"ottingen, Germany. \and 
    \email{avivofir@astro.physik.uni-goettingen.de}}

    \date{Received XXX; accepted YYY}


   \abstract
    {The primary goal of the \emph{Kepler} mission is the measurement of the frequency of Earth-like planets around Sun-like stars. However, the confirmation of the smallest of \emph{Kepler}'s candidates in long periods around FGK dwarfs is extremely difficult or even beyond the limit of current radial velocity technology. Transit timing variations (TTVs) may offer the possibility for such confirmations of near-resonant multiple systems by the mutual gravitational interaction of the planets.}
    {We previously detected the second planet candidate in the KOI\,1574 system. The two candidates are relatively long-period (about 114d and 191d) and in 5:3 resonance. We therefore search for TTVs in this particularly promising system.}
    {The full \emph{Kepler} data was detrended with the proven SARS pipeline. The entire data allowed searching for TTVs of the above signals, as well as searching for additional transit-like signals.}
    {We detect strong anti-correlated TTVs of the 114d and 191d signals, dynamically confirming them as members of the same system. Dynamical simulations reproducing the observed TTVs allow us to also determine the masses of the planets. KOI 1574.01 (hereafter Kepler-87\,b) was found to have a radius of \RpIn $\pm$ \RpInErr $R_\oplus$ and a mass of {\bf \MpIn $\pm$ \MpInErr $M_\oplus$}, and KOI 1574.02 (Kepler-87\,c) was found to have a radius of \RpOut $\pm$ \RpOutErr $R_\oplus$ and a mass of {\bf \MpOut $\pm$ \MpOutErr $M_\oplus$}. Both planets have low densities of \RhoIn and \RhoOut $g\,cm^{-3}$, respectively, which is non-trivial for such cold and old (7-8 Gyr) planets. Specifically, Kepler-87\,c is the lowest-density planet in the super-Earth mass range.
    Both planets are thus particularly amenable to modeling and planetary structure studies, and also present an interesting case were ground-based photometric follow-up of {\emph Kepler} planets is very desirable. Finally, we also detect two more short-period super-Earth sized planetary ($<2R_\oplus$) candidates in the system, making the relatively high multiplicity of this system notable against the general paucity of multiple systems in the presence of giant planets like Kepler-87\,b.}
    {}

\titlerunning{A low-density super-Earth mass planet around Kepler-87}

\maketitle
%

\section{Introduction}

It is very difficult to detect, and even more difficult to confirm, small planets orbiting in long periods around their host star -- where the liquid-water habitable zone (HZ) resides. This is relatively easier for M dwarf host stars since they are both smaller and lighter than Sun-like stars, making the respective transit and radial velocity signals larger. These considerations, coupled with the M dwarfs prevalence in the stellar population, are behind the great interest in M dwarfs and their HZ planets (e.g. Anglada-Escud\'e \etal 2012). However, for more massive stars, like the Sun and the bulk of the \emph{Kepler} target stars, small HZ planets remain elusive targets. The few small HZ planets that were detected so far (e.g. Kepler-22, Borucki \etal 2012) are all either around M dwarfs or with no dynamical confirmation (i.e., no mass measured). One way to positively detect such objects is using transit timing variations: in near-resonant systems the amplitude of these variations can allow detecting small planets, even down to Earth-mass, in {\emph Kepler}'s data (Holman \& Murray 2005).

KOI\,1574 was flagged in Batalha \etal 2013 as having a relatively deep ($\sim 0.5\%$) candidate with a period of $P_{01}\approx114$d (hereafter KOI\,1574.01). Ofir \& Dreizler (2013, hereafter OD13) re-analyzed all of {\emph Kepler}'s KOIs and found 84 new transiting planets candidates in these light curves. Among them, OD13 identified an additional candidate in the KOI\,1574 system using quarters 0 through 6 data. The additional outer candidate is in 5:3 resonance with KOI\,1574.01, or a period of $P_{02} \approx 191d$ (hereafter also KOI\,1574.02). In this work we describe the KOI 1574 planetary system, and KOI\,1574.02 in particular, as the first detection of a transiting super-Earth mass in a long period (near the HZ of KOI 1574). We present the spectral analysis of the host star on \S \ref{SpecAnalysis}, light curve processing on \S \ref{LCanalysis}, the observed TTVs and the resultant derived masses on \S \ref{TTVs}, and conclude.

\section{Spectral analysis}
\label{SpecAnalysis}

\subsection{Observed spectra}
We used two spectra of KOI\,1574 for spectral analysis. The first one has been downloaded from the Kepler Community Follow-up Observing Program (CFOP) website \footnote{\texttt{https://cfop.ipac.caltech.edu/home/}}. The spectrum was taken by Erik Brugamyer and William Cochran using the (Tull) Coude spectrograph at the 2.7 meter (Harlan J. Smith) telescope at the McDonald Observatory on JD 2455703.83350 with an exposure time of 2900\,sec. The spectrograph has a resolution of 60\,000 and covers a spectral range from about 3750\,\AA\, up to 10\,000\,\AA. The useful range is however restricted to 4250-9000\,\AA\, due to low signal-to-noise outside this range. The spectrum has been reduced with IRAF applying standard processing.

The second spectrum has been taken with the Hobby Eberly Telescope (HET; Ramsey et al., 1998) with the High Resolution Spectrograph (HRS; Tull et al., 1995) in a setup (15k central 600g5271 2as 2sky IS0 GC0 2x5) which provides a resolution of 15 000 and a wavelength coverage from 4260\,\AA\ to 6220\,\AA. It has been obtained on 2012-09-23 ($\mathrm{JD}=2456193.705981$) with an exposure time of 2400 sec. Using the IDL based {\tt REDUCE} package (Piskunov \& Valenti, 2002) the spectrum was bias corrected, flat-fielded, optimally extracted, and finally wavelength calibrated using a ThAr lamp.

\subsection{Model atmosphere fitting}

At the Kepler-CFOP web page Sam Quinn provided the following analysis for the McDonald coude spectrum: T$_{\rm eff}$=5750\,K, $\log(g)=4.0$, V$_{\rm rot}$=4\,km/s at solar metallicity. It has to be noted that the uncertainties are estimated to be $\pm$125\,K, $\pm$0.25\,dex, and $\pm$1\,km/s for the effective temperature surface gravity and rotational velocity. Due to the correlation between metallicity and effective temperature, a variation of the metallicity by 0.2 dex would result in an additional uncertainty of the the effective temperature of about 200\,K. From a comparison of these values with stellar evolution models, the mass determination of the central star would be uncertain by 20-30\%.

In order to improve the parameter determination we performed a model atmosphere fitting using the newest PHOENIX model grid (Husser \etal 2012). Models are available in steps of 100\,K, 0.5\,dex, and 0.5\,dex in effective temperature, surface gravity and metallicity. The microturbulence is not a free parameter but is derived from a scaling law using the mean convective velocity with each model. This scaling relation has been calibrated using 3D radiative transport on 3D hydrodynamical simulations. Other improvements compared to earlier models is a new equation of state as well as spherical symmetry for all models. We use a Levenberg-Marquardt optimization to fit the effective temperature, surface gravity, metallicity, and rotational velocity (only in the higher resolved McDonald spectrum) simultaneously with a polynomial for the continuum for each spectral order. The surface gravity, however, was allowed to vary in a small interval log(g)=[3.9,4.0] only, which can be derived from the ratio of the stellar radius and the orbital period of the planet taken from the light curve analysis (\S \ref{LCanalysis}), Kepler's Third Law, and stellar evolution models. The final values and errors are weighted means over all spectral orders. It should be noted that we multiplied the errors by a factor of two in order to account for systematic errors, e.g. from the fact that the model atmospheres have to be calculated in 1D, allowing to treat convection only in the mixing length approximation, or from the fact that the instrumental broadening was approximated with a Gauss-profile. 

\begin{figure}[tbp]\includegraphics[trim=0 350 0 150, width=0.5\textwidth]{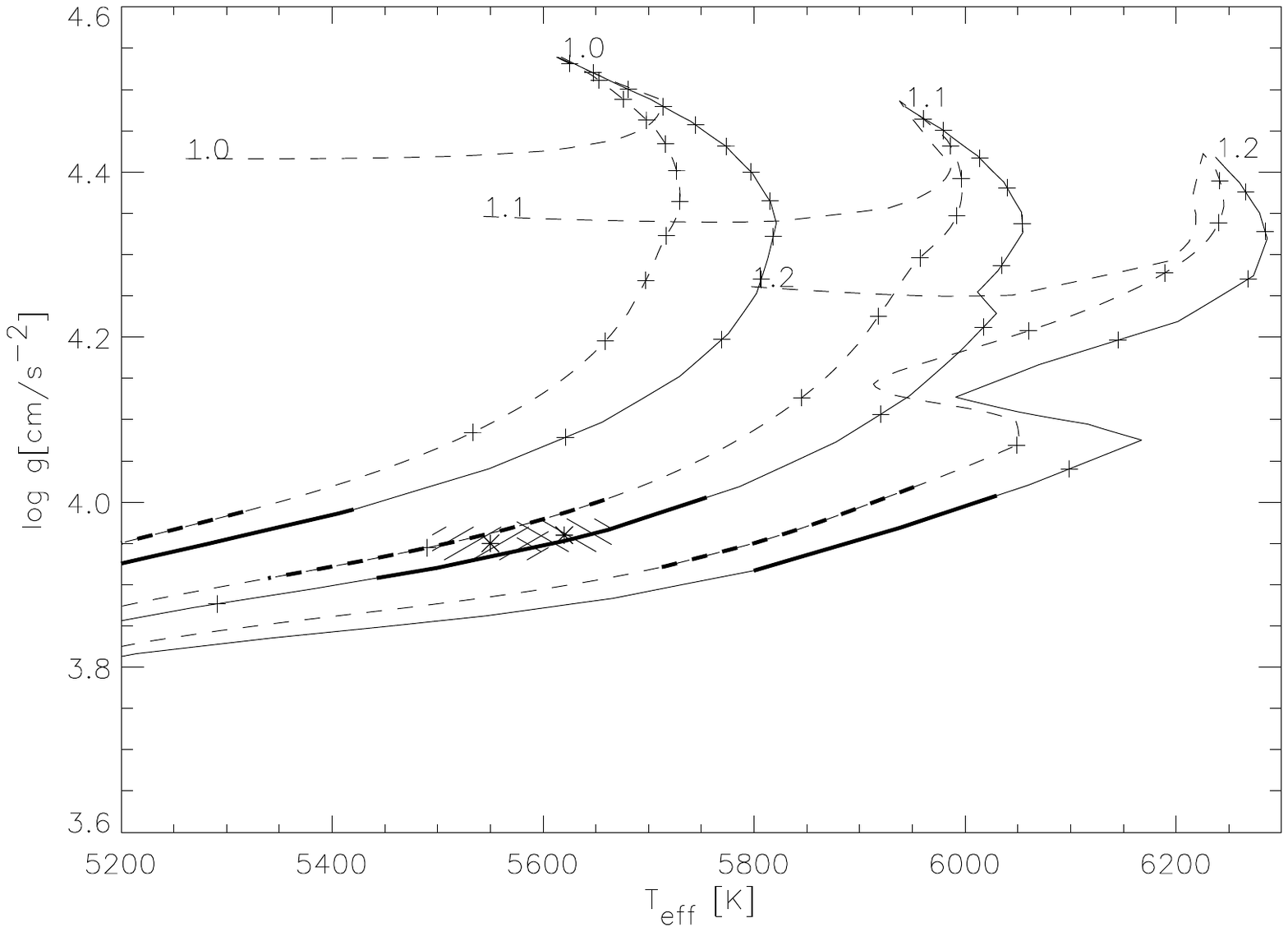}
\caption{Stellar parameters from Table\,\ref{Tabspectro} (shaded area) compared to Padova stellar evolution models (solid line) and Y$^2$ models (dashed line). Note that the models differ slightly in their chemical composition. The bold regions on the tracks indicate the ratio of the stellar radius to the semi major axis of the planetary orbit derived from its orbital period, the mass of the stellar model, and Kepler's Third Law. The ``+'' symbols indicate 1Gyr time steps.}
\label{Figspectro}
\end{figure}

The stellar parameters determined from the two spectra (see Table\,\ref{Tabspectro} and also Fig.\,\ref{Figspectro})  marginally agree within their 1\,$\sigma$ errors. For the final stellar parameters we adopt a mean from the two determinations. The parameters reported in CFOP reveal a slightly higher effective temperature. It should be noted that a solar abundance was assumed in that case. With our slightly sub-solar metallicity a somewhat lower effective temperature is needed to achieve similar line strengths of the mainly neutral metal lines. 

We use the stellar parameters to compare KOI\,1574 with Padova stellar evolution models (mass fraction for hydrogen $X=72.3\%$, Helium 26\% and metals $Z=1.7\%$; (Bertelli \etal 2008)) as well as with Y$^2$ models (mass fraction for hydrogen $X=71\%$, Helium 27\% and metals $Z=2\%$ (Yi \etal 2001, Kim \etal 2002, Yi, Kim, \& Demarque 2003, Demarque \etal 2004)). KOI\,1574 is a star at the end of its main sequence phase. The slightly sub-solar metallicity is consistent with an age of about 7-8\,Gyr. We derive a stellar mass of 1.1\,M$_\odot \pm 0.05$\,M$_\odot$, which takes into account the dependence of the chemical composition as well as possible systematic errors in the evolution models. This results in a stellar radius of 1.82\,R$_\odot \pm 0.04$\,R$_\odot$.

\begin{table}
\begin{tabular}{llll}
\hline
                        & McDonald          & HET & mean \\
\hline
Resolution              & 60\,000           & 15\,000 & \\
\hline
T$_{\rm eff}$ [K]         & 5550$\,\pm$ 60    & 5640$\,\pm$ 45   & 5600$\,\pm$ 50\\  
log(g) [cgs]              & 3.95$\,\pm$ 0.02  & 3.96$\,\pm$ 0.02 & 3.96$\,\pm$ 0.02\\
\rule{0cm}{0.01mm}[Fe/H]  & -0.18$\,\pm$ 0.04 &-0.16$\,\pm$ 0.03 & -0.17$\,\pm$ 0.03\\
 V$_{\rm rot}$ [km/s]     & 4.3$\,\pm$ 0.2    & fixed            & 4.3$\,\pm$ 0.2\\
\hline
\end{tabular}
\caption{Stellar parameters derived from our model atmosphere fit for the McDonald 2.7m spectrum and the HET High Resolution Spectrograph. It should be noted that the surface gravity was allowed to vary in a small interval determined by the stellar density derived from the light curve fits only.}
\label{Tabspectro}
\end{table}

\section{Light curve analysis}
\label{LCanalysis}

\subsection{Preprocessing and detection}
\label{Preprocessing}

We use the full Kepler data -- quarters 1 through 16 -- in our analysis. The additional data includes a few more transits of KOI 1574.01 relative to OD13, and importantly - the third and forth transit events of KOI 1574.02 (unfortunately the star fall on the inactive Module 3 during Quarters 7, 11 and 15 and transits that did occur were not observed). We applied nearly the same processing as in OD13 to the entire dataset. In short, it includes the removal of long-term trends by the application of a median filter to each continuous section individually, and the identification and removal of both additive and relative systematic effects in the data simultaneously with the SARS algorithm (Ofir \etal 2010). The only differences were: (a) the inclusion of a correction for crowding and flux ratio effects as in eq. 2 of Stumpe \etal (2012), (b) the use of the newly-available target-specific status indicator to identify continuous sections instead of a global anomaly table, and (c) active avoidance of variable stars (KOIs, eclipsing binaries and red giants) from the SARS learning set.

We re-searched the KOI\,1574 system for transit signals and found the previous two signals very significantly. We then searched for transit time variations (TTVs) for each signal by fitted the linear-ephemeris model (computed using the Mandel and Agol (2002) formalism) to each one of the individual transits, allowing only for the time of mid-transit to vary. Indeed, KOIs 1574.01 and 1574.02 showe strongly anti-correlated TTVs (see Fig. \ref{Figttv} and discussion on \S \ref{TTVs}). These anti-correlated TTVs, coupled with the dynamical simulations that give strong limits on the masses of the two objects, allow us to dynamically confirm the KOIs 1574.01 and 1574.02 signals as true planets in the same planetary system. We therefore name these planets Kepler-87\,b and c, respectively.

After the above initial TTVs-corrected modeling we modeled-out the planets and re-calculated the background long-term trends and re-fitted the planets -- till convergence. We then removed the Kepler-87\,b and c models completely and applied the Optimal Box Least Squares technique (Ofir 2013) to search for additional transit signals in the residuals, and detected two additional short-period transit-like signals with periods of $P_{03}\approx5.83$d and $P_{04}\approx8.97$d above the $7.1\sigma$ significance threshold -- hereafter KOI\,1574.03 and KOI\,1574.04, respectively. We note that the 1574.03 signal was also identified by the Kepler team in the Q0-Q8 data \footnote{As cataloged at the Exoplanet Archive \texttt{ http://exoplanetarchive.ipac.caltech.edu/index.html}}. The new signals also passed all the other tests described in OD13. Actually, At this point we custom-fitted the long-term filter for this particular object: we changed the general segmented median filter to a segmented Savitzky-Golay filter (Savitzky \& Golay 1964) (of second order, in a two-days window span) with iterative $3 \sigma$ clipping, which is better than a simple median filter, and repeated all the above. We note that KOI\,1574.03 and KOI\,1574.04 show no significant TTVs - but the error bars are quite large for such shallow and short-period candidates.

\subsection{Global fit}

\begin{figure}[tbp]\includegraphics[width=0.5\textwidth]{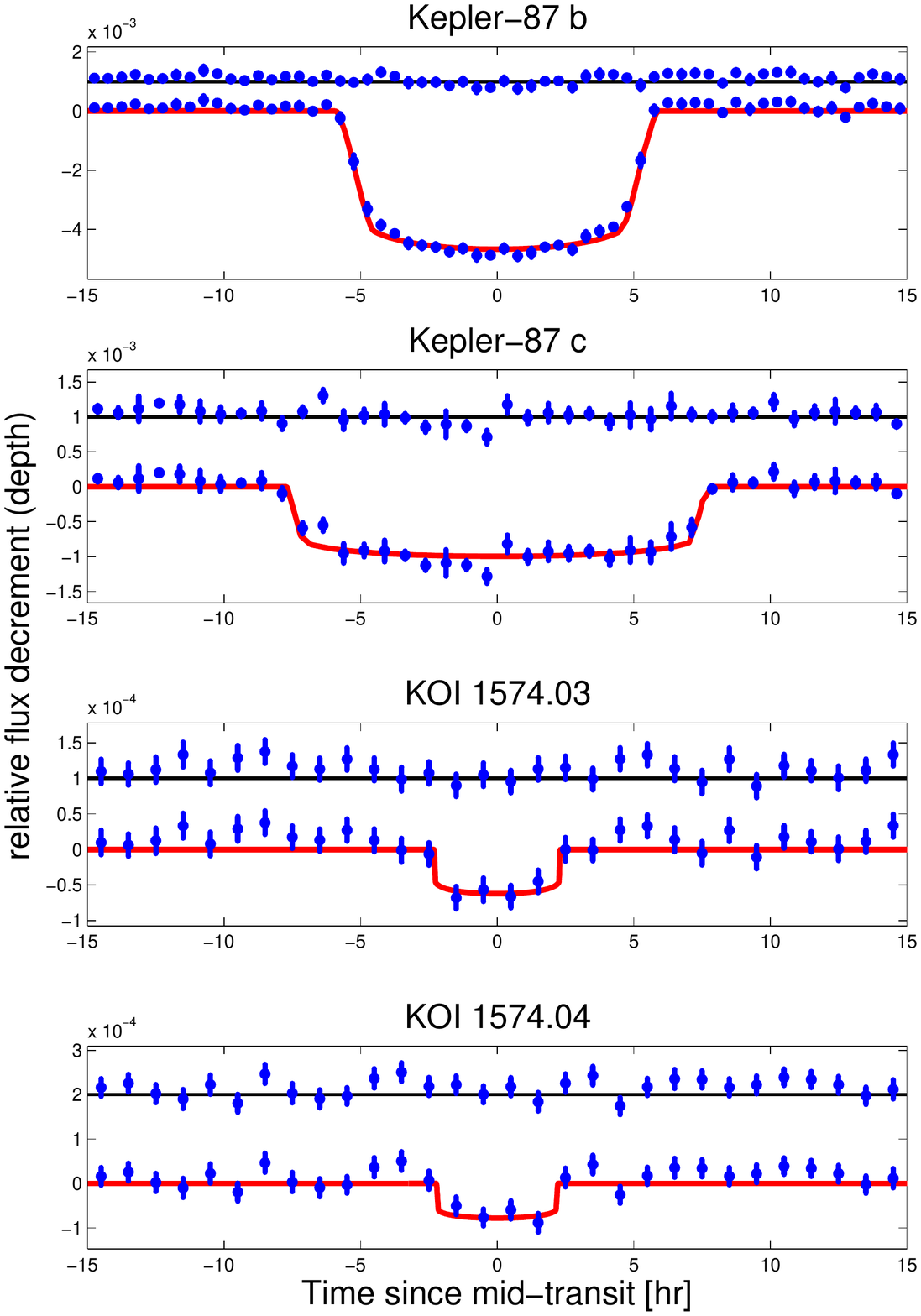}
\caption{Phased and binned (to half-hour bins) light curves of the Kepler-87 system components relative to the time of mid transit, with over plotted best-fitting models. From top to bottom: transiting exoplanets Kepler-87\,b and c and planetary candidates KOIs 1574.03 and 1574.04. Above each light curve we show the model residuals, shifted for clarity. Note the vertical scale in each panel may be different.}
\label{LCs}
\end{figure}

We derived the final system parameters (given in Tables \ref{TimingFit} and \ref{SystemParams}, illustrated in Fig. \ref{LCs}) by simultaneously fitting all four signals. We iterated the procedure below and the background long-term trends fitting several times till convergence, and report the final iteration here. We used circular orbits for all signals, but we allowed the phase of all Kepler-87\,b \& c transits to be set relative to the closest time of mid-transit, and optimized for these times as well. This perturbed-circular fit is valid in the small TTVs regime only, and indeed the largest TTVs detected are about $7.5 \cdot 10^{-5}$ of $P_{01}$ and $6 \cdot 10^{-4}$ of $P_{02}$. The scaled semi-major axis $a/R_*$ parameter was common to all candidates: as in OD13 we scaled it by Kepler's third law for each candidate. The final fit therefore included these parameters: one $a/R_*$, four planet radii $r_p/R_*$, four impact parameters $b_p/R_*$, eleven $T_{mid}$ for Kepler-87\,b, four $T_{mid}$ for Kepler-87\,c, and two linear parameters ($P$ and $T_{mid}$) for each of KOIs 1574.03 and 1574.04 - a total of 28 floating parameters. Once initial results suggested the proximity of Kepler-87\,c to the HZ (below), and since the relevant parameter $a/R_*$ is usually both the most difficult to fit (has the largest error) and may have some sensitivity to the initial starting point, we ran twelve $5\cdot10^5$-step MCMC fits that allowed all the variables to float -- each with a different $a/R_*$ starting points evenly sampled between half and twice our initial estimate. We then checked that all twelve parameters sets converges on consistent values to $1 \sigma$ on all parameters. The total of the 12 chains exhibited a smooth distribution of values up to $\Delta \chi ^2 < 100$ (relative to the global minimum) so we considered as ``burn-in'' of each MCMC chain as all the steps before the first time $\Delta \chi ^2 < 100$ was reached, relative to the global minimum, and concated all these truncated chains (as in Tegmark \etal 2004) to one very large, nearly $6 \cdot 10^6$ steps long, chain that was used for parameter estimation. The final linear ephemeris for Kepler-87\,b and c and their error bars were computed from the distribution of fits to the different $T_{mid}$ along the MCMC chain. We note that the final $a_{01}/R_*$ is smaller than the linear one (given in OD13), as expected: the linear ephemeris fit caused the average signal to appear smeared, and therefore with artificially higher $a_{01}/R_*$.

The equilibrium temperature of exoplanet Kepler-87\,c is mostly derived from the above model's $a/R_*$ axis and the host star $T_{\rm eff}$ using:

\begin{equation}
 T_{eq}=T_{\rm eff}\left(\frac{R_*}{2a}\right)^{1/2}[f(1-A_B)]^{1/4}
\end{equation}

However, the planetary atmospheric parameters flux redistribution factor $f$ and Bond albedo $A_B$ are completely unknown, and have a large effect on the resultant $T_{eq}$.
``Conventional'' values would assume efficient redistribution of the stellar flux ($f=1$) and Earth-like albedo ($A_B$=0.3), and these lead to $T_{eq,02}$=\TeqOut, or just hotter than the inner edge of the habitable zone. However, $f$ and $A_B$ are completely unknown and can vary considerably. Such changes in any of $f$ or $A_B$ can lower the $T_{eq,02}$ to well within the HZ. 


\begin{table}
\caption{Timing results of the perturbed circular orbit fit for the Kepler-87\,b and Kepler-87\,c planets. Residuals are calculated relative to the best-fit linear solution.}
\begin{tabular}{l l l l}
\hline
Best-fit time of& Median time of  & Residuals from   & $1 \sigma$ \\
mid transit	& mid transit	  & linear ephemeris & error      \\
(BJD-2454833)   & (BJD-2454833)   & [d]		     & [d]     	  \\
\hline
\multicolumn{4}{l}{Kepler-87\,b} \\
\hline
165.1564 & 165.1576 & 0.0056 & $_{-0.0024}^{+0.0025}$\\
279.8881 & 279.8884 & 0.0001 & $\pm 0.0017$\\
394.6166 & 394.6161 & -0.0086 & $\pm 0.0016$\\
509.3531 & 509.3527 & -0.0083 & $\pm 0.0014$\\
624.1031 & 624.1034 & 0.0060 & $\pm 0.0014$\\
738.8414 & 738.8409 & 0.0071 & $_{-0.0015}^{+0.0014}$\\
853.5690 & 853.5700 & -0.0001 & $_{-0.0017}^{+0.0018}$\\
968.3024 & 968.3010 & -0.0055 & $\pm 0.0020$\\
1197.7846 & 1197.7845 & 0.0053 & $\pm 0.0015$\\
1312.5218 & 1312.5223 & 0.0068 & $_{-0.0015}^{+0.0016}$\\
1541.9800 & 1541.9798 & -0.0085 & $_{-0.0017}^{+0.0018}$\\
\hline
\multicolumn{4}{l}{Kepler-87\,c} \\
\hline
286.0894 & 286.0900 & -0.1606 & $_{-0.0069}^{+0.0059}$\\
477.6302 & 477.6292 & 0.1467 & $_{-0.0053}^{+0.0051}$\\
860.0537 & 860.0483 & 0.1022 & $_{-0.0065}^{+0.0058}$\\
1242.3155 & 1242.3152 & -0.0945 & $_{-0.0051}^{+0.0052}$\\
\hline
\end{tabular}
\label{TimingFit}
\end{table}

\begin{table}
\caption{Predicted transit times for the Kepler-87\,b and c planets for the next few years of the adopted model.}
\begin{tabular}{l l l l|l l}
\hline
\multicolumn{6}{c}{Predicted time of mid transit (BJD-2454833)}\\
& \multicolumn{2}{c}{Kepler-87\,b} & & \multicolumn{2}{c}{Kepler-87\,c} \\
\hline
1312.521$\!\!$  & $\!\!$1771.462$\!\!$ & $\!\!$2230.386$\!\!$ & $\!\!$2574.589$\!\!$ & $\!\!$1434.256$\!\!$ & 2200.750\\
1427.251$\!\!$  & $\!\!$1886.196$\!\!$ & $\!\!$2345.130$\!\!$ & $\!\!$2689.316$\!\!$ & $\!\!$1625.957$\!\!$ & 2391.878\\
1541.980$\!\!$  & $\!\!$2000.924$\!\!$ & $\!\!$2459.862$\!\!$ & $\!\!$2804.046$\!\!$ & $\!\!$1816.789$\!\!$ & 2584.208\\
1656.714$\!\!$  & $\!\!$2115.653$\!\!$ & $\!\!$2459.862$\!\!$ & $\!\!$3033.517$\!\!$ & $\!\!$2008.904$\!\!$ & 2776.229\\
\hline
\end{tabular}
\label{Preditions}
\end{table}

\begin{table}
\caption{Observed and derived parameters for the Kepler-87 (KOI\,1574) system. Subscript XX is understood to be the relevant parameter for signal 1574.XX. All times are BJD-2454833.}
\begin{tabular}{l l l l}
\hline
\multicolumn{4}{l}{Observed parameters} \\
Quantity     &	Best Fit     &  median &  1 $\sigma$ error  \\
\hline
$LinearP_{01}$[d] &  114.73635 &  114.73631 & $\pm    0.00015$\\
$a_{01}/R_*   $ &       57.4 &       56.8 & $_{      -1.2}^{+       1.4}$\\
$r_{01}/R_*   $ &    0.06855 &    0.06859 & $_{  -0.00028}^{+   0.00026}$\\
$b_{01}/R_*   $ &      0.727 &      0.732 & $_{    -0.017}^{+     0.014}$\\
$LinearP_{02}$[d] &   191.2318 &   191.2315 & $\pm     0.0015$\\
$a_{02}       $ &       80.7 &       79.9 & $_{      -1.7}^{+       2.0}$\\
$r_{02}/R_*   $ &    0.03123 &    0.03119 & $_{  -0.00042}^{+   0.00041}$\\
$b_{02}/R_*   $ &      0.579 &      0.591 & $_{    -0.031}^{+     0.025}$\\
$P_{03}$ [d]    &   5.833904 &   5.833902 & $_{ -0.000059}^{+  0.000050}$\\
$T_{mid,3}$\,$^{(a)}$ &   517.6762 &   517.6753 & $_{   -0.0044}^{+    0.0047}$\\
$a_{03}       $ &       7.88 &       7.80 & $_{     -0.17}^{+      0.19}$\\
$r_{03}/R_*   $ &    0.00856 &    0.00853 & $_{  -0.00043}^{+   0.00040}$\\
$b_{03}/R_*   $ &      0.567 &      0.591 & $_{    -0.068}^{+     0.052}$\\
$P_{04}$ [d]    &    8.97741 &    8.97730 & $\pm    0.00011$\\
$a_{04}       $ &      10.50 &      10.40 & $_{     -0.22}^{+      0.26}$\\
$T_{mid,4}$ &   519.4697 &   519.4679 & $_{   -0.0044}^{+    0.0047}$\\
$r_{04}/R_*   $ &    0.00983 &    0.00925 & $_{  -0.00049}^{+   0.00047}$\\
$b_{04}/R_*   $ &      0.746 &      0.742 & $_{    -0.038}^{+     0.029}$\\

\hline
\multicolumn{4}{l}{Stellar parameters derived from spectroscopy} \\
\hline
$M_*$ [$M_\odot$]	&   1.1	   	& & $\pm$ 0.05 \\
$R_*$ [$R_\odot$]	&   1.82	& & $\pm$ 0.04 \\
Age [Gyr]		&   7\,--\,8	& & \\

\hline
\multicolumn{4}{l}{Parameters from dynamical modeling} \\
Quantity     &	mean	&  median &  1 $\sigma$ error  \\
\hline
$M_* $ [M$_\odot$]   &  1.05       &  1.08   & $\pm$ 0.06     \\

$m_{01}$ [M$_\oplus$]  & 324.2       & 326.1   & $\pm$ 8.8      \\
$a_{01t}$ [AU]         & 0.471       & 0.474   & $\pm$ 0.010    \\
$P_{01}$ [d]           & 114.7309     & 114.7310 & $\pm$ 0.0005    \\
$e_{01}$               & 0.036       & 0.039   & $\pm$ 0.009    \\
$\omega_{01} [^\circ]$ & 238.6       &  255.3   & $\pm$ 27.6     \\
M$_{01} [^\circ]$      & 293.0       & 296.7   & $\pm$ 23.6     \\

$m_{02}$ [M$_\oplus$]  &   6.4       &   6.5   & $\pm$ 0.8      \\
$a_{02}$ [AU]          & 0.664       & 0.668   & $\pm$ 0.013    \\
$P_{02}$ [d]           & 192.363     & 192.389 & $\pm$ 0.074    \\
$e_{02}$               & 0.039       & 0.042   & $\pm$ 0.012    \\
$\omega_{02} [^\circ]$ & 223.2       &  240.1  & $\pm$ 18.8     \\
M$_{02} [^\circ]$      & 291.4       & 297.3   & $\pm$ 14.4     \\

\hline
\multicolumn{4}{l}{Derived physical parameters} \\
Quantity     &	Best Fit &  median &  1 $\sigma$ error  \\
\hline
$r_{01}\,\, [R_\oplus]$  &      \RpIn &      13.49 & $\pm       \RpInErr $\\
$a_{01}\,\, [\mathrm{AU}] ^{(b)}$ &      0.481 &      0.476 & $_{    -0.028}^{+     0.026}$\\
$i_{01}\,\, [^\circ]$    &     89.274 &     89.262 & $_{    -0.030}^{+     0.034}$\\
$\rho_{01}\,\, [g \,\, cm^{-3}]$      &     \RhoIn &      0.728 & $\pm      \RhoInErr$\\
$T_{\mathrm{eq,01}}\,\, [^\circ K]$   &     \TeqIn &      480.5 & $\pm        4.3$\\

$r_{02}\,\, [R_\oplus]$  &     \RpOut &       6.14 & $\pm       \RpOutErr $\\
$a_{02}\,\, [\mathrm{AU}] ^{(b)}$ &      0.676 &      0.669 & $_{    -0.040}^{+     0.037}$\\
$i_{02}\,\, [^\circ]$    &     89.588 &     89.576 & $_{    -0.027}^{+     0.031}$\\
$\rho_{02}\,\, [g \,\, cm^{-3}]$      &    \RhoOut &      0.153 & $\pm      \RhoOutErr$\\
$T_{\mathrm{eq,02}}\,\, [^\circ K]$   &    \TeqOut &      405.2 & $\pm        3.6$\\

$r_{03}\,\, [R_\oplus]$  &     \RpIII &       1.68 & $\pm       \RpIIIErr $\\
$a_{03}\,\, [\mathrm{AU}] ^{(b)}$ &     0.0660 &     0.0654 & $_{   -0.0039}^{+    0.0036}$\\
$i_{03}\,\, [^\circ]$    &      85.87 &      85.66 & $_{     -0.45}^{+      0.57}$\\
$T_{\mathrm{eq,03}}\,\, [^\circ K]$   &       1291 &       1297 & $\pm         12$\\

$r_{04}\,\, [R_\oplus]$  &      \RpIV &       1.82 & $\pm       \RpIVErr $\\
$a_{04}\,\, [\mathrm{AU}] ^{(b)}$ &     0.0880 &     0.0871 & $_{   -0.0052}^{+    0.0048}$\\
$i_{04}\,\, [^\circ]$    &      85.93 &      85.91 & $_{     -0.22}^{+      0.27}$\\
$T_{\mathrm{eq,04}}\,\, [^\circ K]$   &     1117.9 &     1123.3 & $\pm       10.0$\\

\hline
\multicolumn{4}{l}{\footnotesize{notes:}} \\
\multicolumn{4}{l}{\footnotesize{(a) A secondary solution, disfavored by $\Delta\chi^2\approx5$ is 517.6524}} \\
\multicolumn{4}{l}{\footnotesize{(b) calculated using $R_*$ times $a_{XX}/R_*$, not from dynamics}} \\

\end{tabular}
\label{SystemParams}
\end{table}

\section{Transit timing variations}
\label{TTVs}

\begin{figure}[tbp]\includegraphics[trim=0 350 0 100, width=0.5\textwidth]{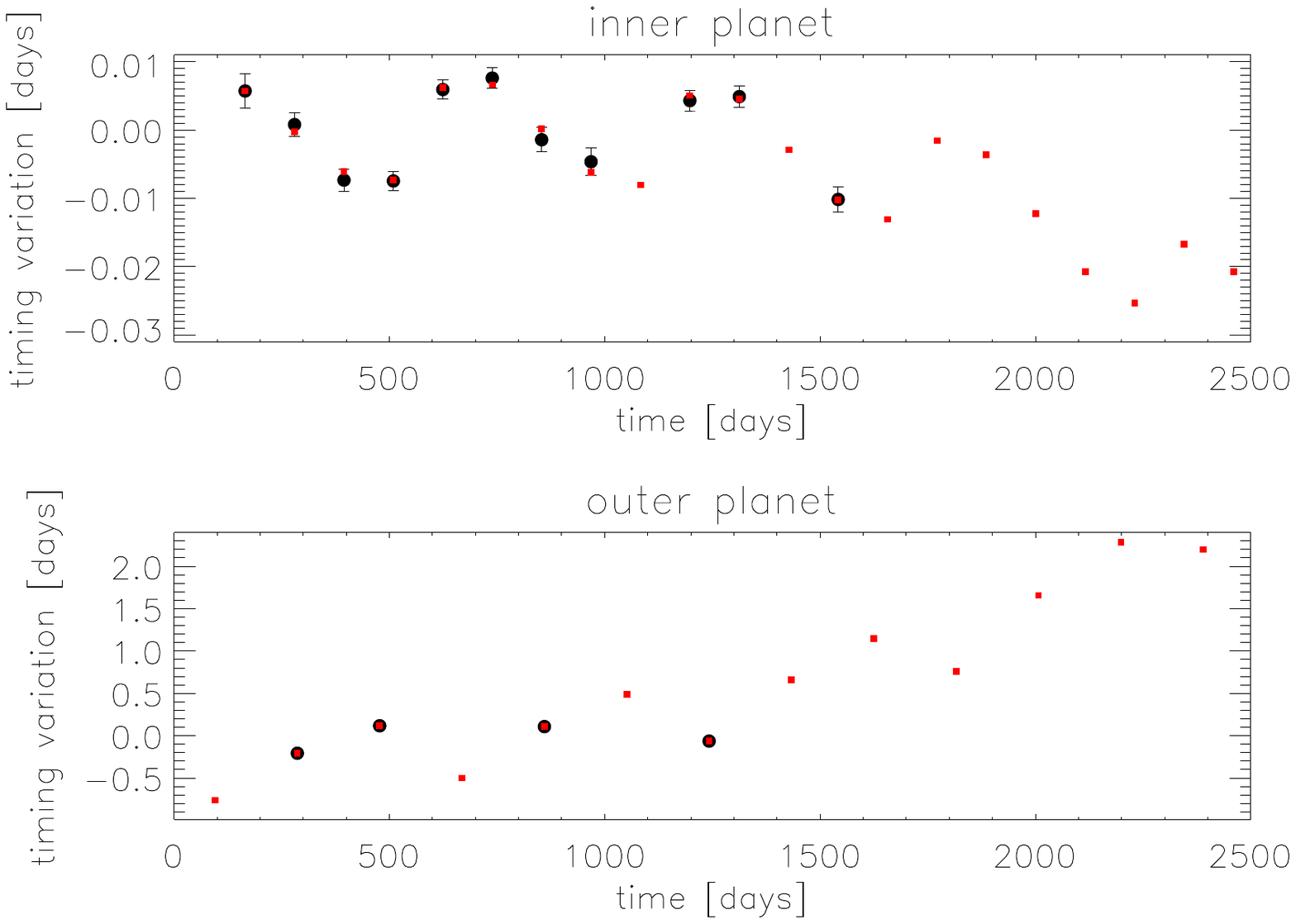}
\caption{Observed transit timing variations relative to a linear ephemeris (black) over plotted with the best model closest to the parameters of Table\,\ref{SystemParams} (red squares). Top and bottom panels are for Kepler-87\,b and c, respectively. The error bars for Kepler-87\,c are smaller than the size of the symbols.}
\label{Figttv}
\end{figure}

Fig. \ref{Figttv} shows the observed TTVs for Kepler-87\,b and c with their error bars (relative to linear ephemeris). A search for the best-fitting sine for the TTVs of Kepler-87\,b (which has enough data points to perform such an analysis), gave a single peak with a super-period of $514_{-80}^{+38}$d, which agrees with the expected $5 \times P_{01}=573.69$ to $1.5\sigma$, further showing that the TTVs are indeed caused by the interaction between Kepler-87\,b and c. Finally, one can prove that these TTVs are indeed from planet-planet interaction by looking for anti-correlation between the TTVs of Kepler-87\,b and c (Ford \etal 2012, Steffen \etal 2012 and Fabrycky \etal 2012), and indeed they exists, and this means that: (a) Kepler-87\,b and Kepler-87\,c are indeed interacting planets in the same system, and (b) we can try to determine the masses of both planets.

We used the hybrid symplectic integrator within the {\it Mercury} package (Chambers 1999), which we have run with a constant time step of 0.5 days, i.e. less than 1\% of the orbital period of the $P_{01}$ planet. We assume co-planar orbits, well justified from the transit fitting (see Table\,\ref{SystemParams}), which together with the stellar and the two planetary masses result in 11 free parameters for the 15 measured transit timings. The two inner planet candidates KOI\,1574.03 and KOI\,1574.04 have not been taken into account for the dynamical analysis.

\begin{figure}[tbp]\includegraphics[trim=0 350 0 100, width=0.5\textwidth]{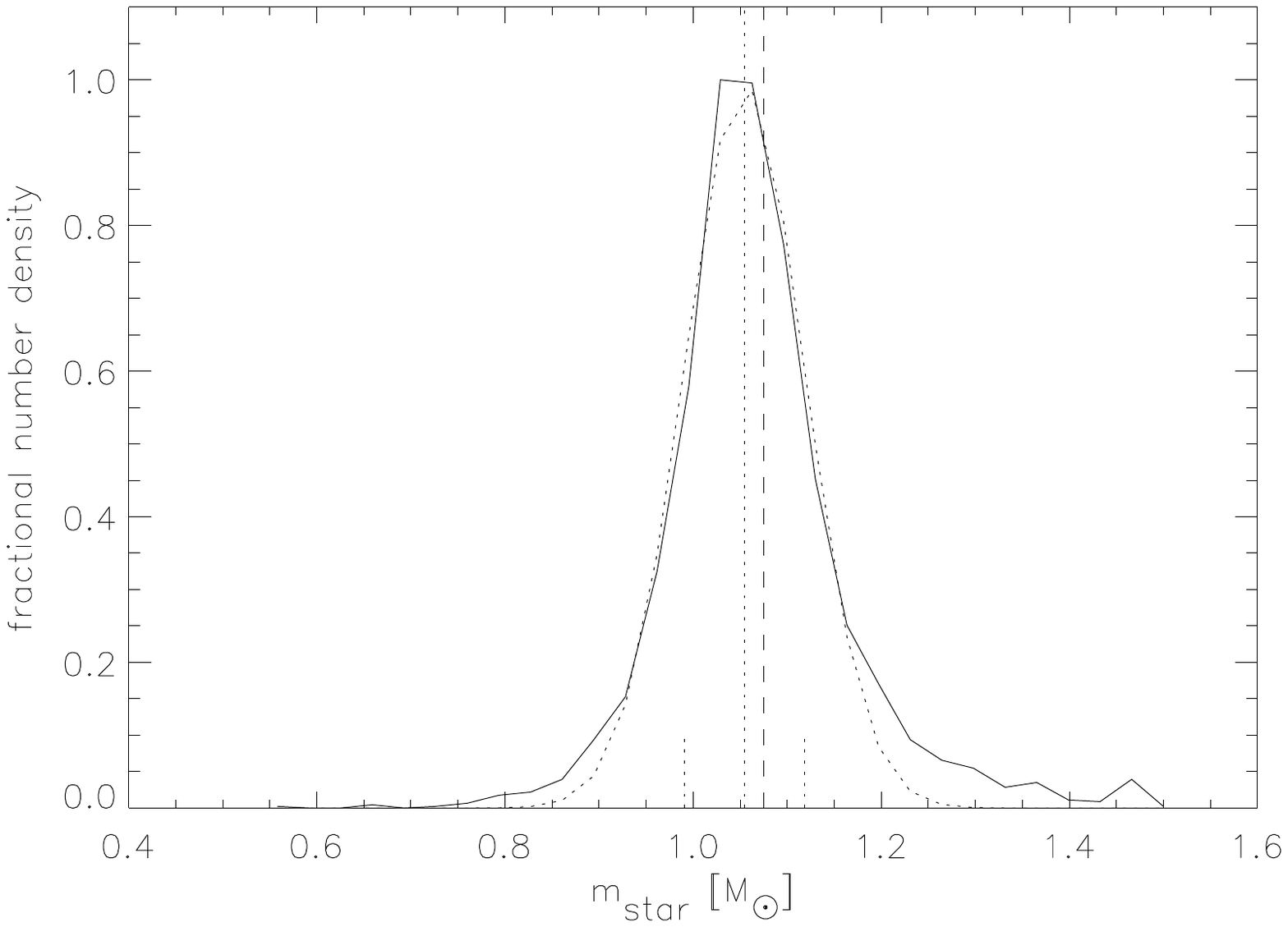}
\caption{Histogram of the stellar mass obtained from
  Levenberg-Marquardt fits starting at 2500 random initial values (full line) fitted with a Gaussian (dotted line). The mean (dotted) and the median (dashed) of the distribution are indicated as long vertical lines, the
  $1\,\sigma$ error as short vertical lines.} 
\label{Figdyn}
\end{figure}

\begin{figure}[tbp]\includegraphics[trim=0 350 0 100, width=0.5\textwidth]{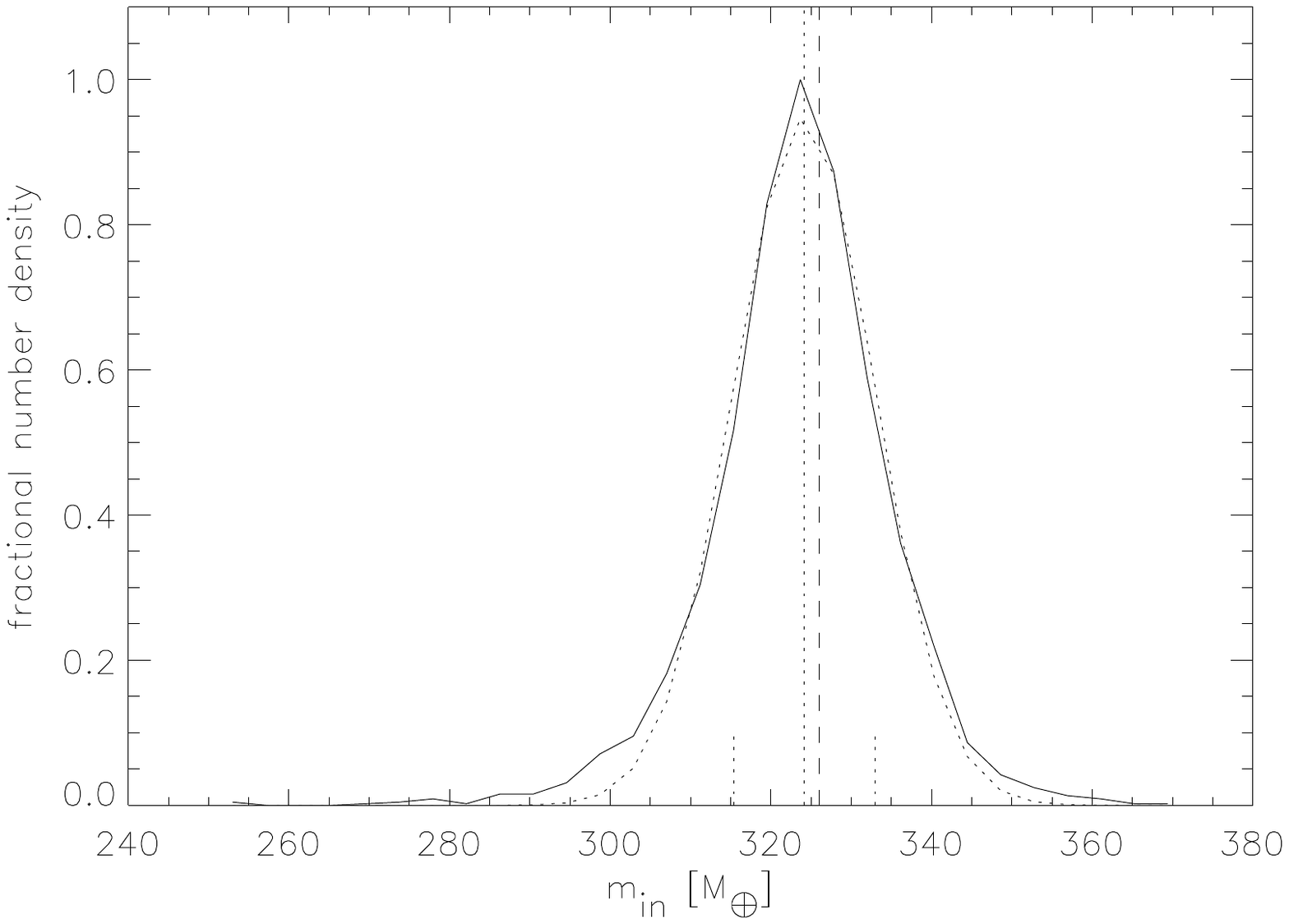}
\caption{Histogram of the mass of the inner planet obtained from
  Levenberg-Marquardt fits starting at 2500 random initial values (full line) fitted with a Gaussian (dotted line). The mean (dotted) and the median (dashed) of the distribution are indicated as long vertical lines, the $1\,\sigma$ error as short vertical lines.} 
\label{Figdyn_m_in}
\end{figure}

\begin{figure}[tbp]\includegraphics[trim=0 350 0 100, width=0.5\textwidth]{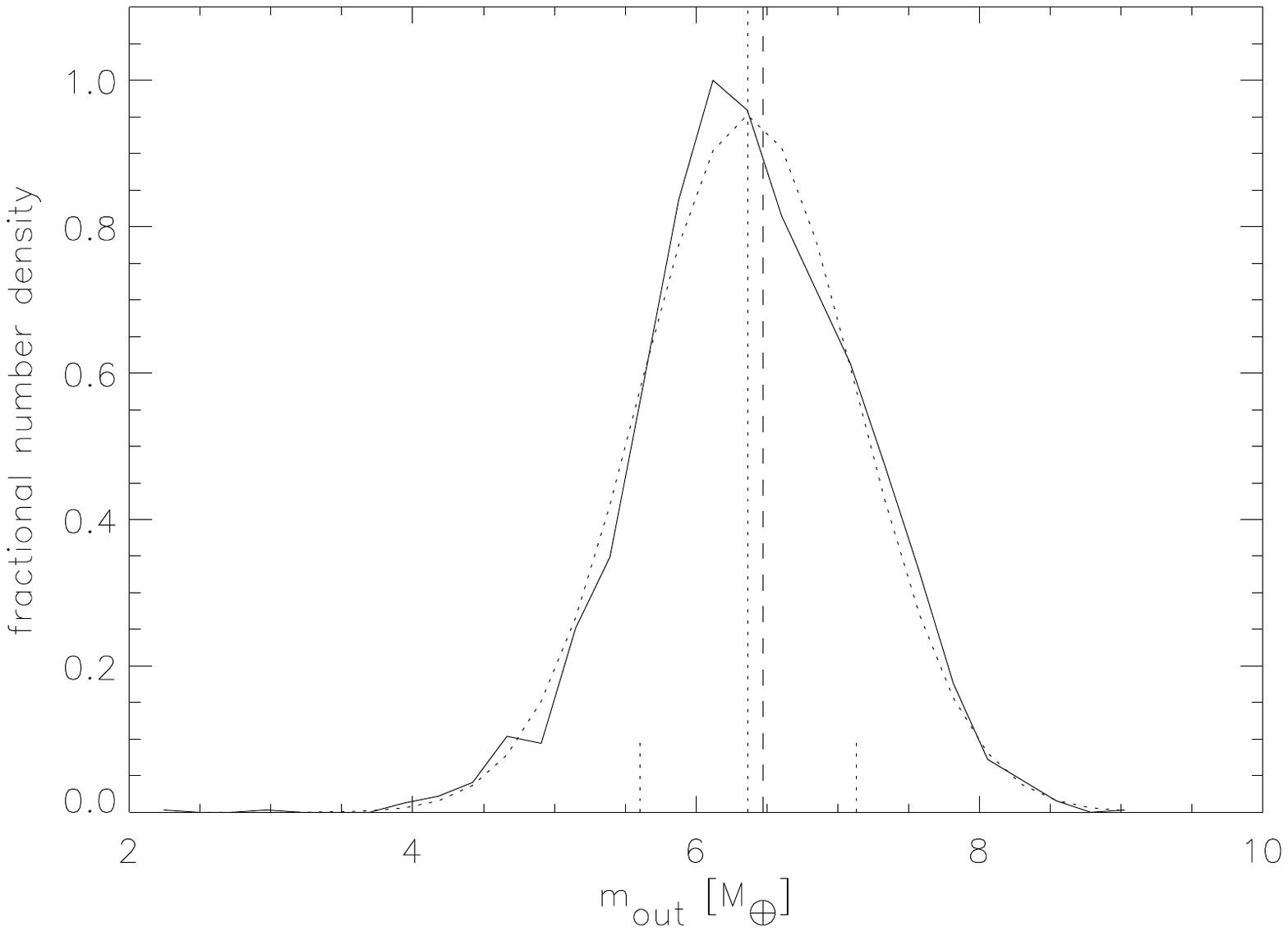}
\caption{Histogram of the mass of the outer planet obtained from
  Levenberg-Marquardt fits starting at 2500 random initial values (full line) fitted with a Gaussian (dotted line). The mean (dotted) and the median (dashed) of the distribution are indicated as long vertical lines, the $1\,\sigma$ error as short vertical lines.} 
\label{Figdyn_m_out}
\end{figure}

Given the stellar mass from the spectral analysis as well as the orbital periods from the light curve, a reasonable set of start parameters can be estimated from the ratio of the TTV amplitudes and the phase of the TTV variations. From preliminary stability calculations the eccentricity could also be limited to be less or equal to about 0.1. We use the \textit{IDL} routine \textit{mpfit}, a Levenberg-Marquardt optimization, to fit the calculated transit timing variations as function of the stellar and the two planetary masses, the semi-major axes, eccentricities, lengths of peri-astron, as well as mean anomalies at $t=0$. We then use this fit to generate 2500 random starting values within the error range provided by \textit{mpfit} and converged them as well. From the resulting sample we derive histograms of for the parameters which allows to obtain mean values and uncertainties. Figures\,\ref{Figdyn}, \ref{Figdyn_m_in}, and \ref{Figdyn_m_out} shows the resulting distribution of the stellar and the planetary masses.

We like to note that the mean stellar mass derived from the TTV analysis is close to the spectroscopic mass derived with the stellar density constraint (Sect.\,\ref{SpecAnalysis}). The fit parameters are listed in Table\,\ref{SystemParams}. The mean masses for the two planets are \MpOut $\pm$ \MpOutErr$M_{\rm \oplus}$ and \MpIn $\pm$ \MpInErr$M_{\rm \oplus}$ for the Kepler-87\,c and b planets respectively. The bulk densities of the planets are determined to be $\rho_{01}=$\RhoIn $\pm$ \RhoInErr and be $\rho_{02}=$\RhoOut $\pm$ \RhoOutErr $\,g\,cm^{-3}$, so the inner planet is a Jupiter mass planet with a Saturn-like density while the outer planet is a very low density planet in the super-Earth mass regime. The eccentricities are low, i.e. the $3\,\sigma$ errors are within the stable regime which allows eccentricities below about 0.1. Within the uncertainties, the lengths of peri-astron of the two orbits are aligned. 

While the total number of measurements is larger than the free parameters, the number of TTV measurements for the outer planet (4) is low. The TTVs of the outer planet are mainly constraining by the mass of the inner planet (TTV amplitude), the eccentricity and periastron length of the outer planet (phase shift against TTV of inner planet and shape of TTV), by the mean anomaly of the outer planet (time of first transit) as well as by the mean orbital period of the outer planet. At first, it seems that the problem is over-determined, but these parameters are also constrained by the TTVs of the inner planet, however, more indirectly from the overall dynamical behavior of the three-body system. We therefore also analyzed the TTVs with a restricted set of parameters, i.e. we fixed the eccentricity and the length of periastron of the outer planet to their mean values. This does not change the results.

\begin{figure}[tbp]\includegraphics[trim=0 350 0 100, width=0.5\textwidth]{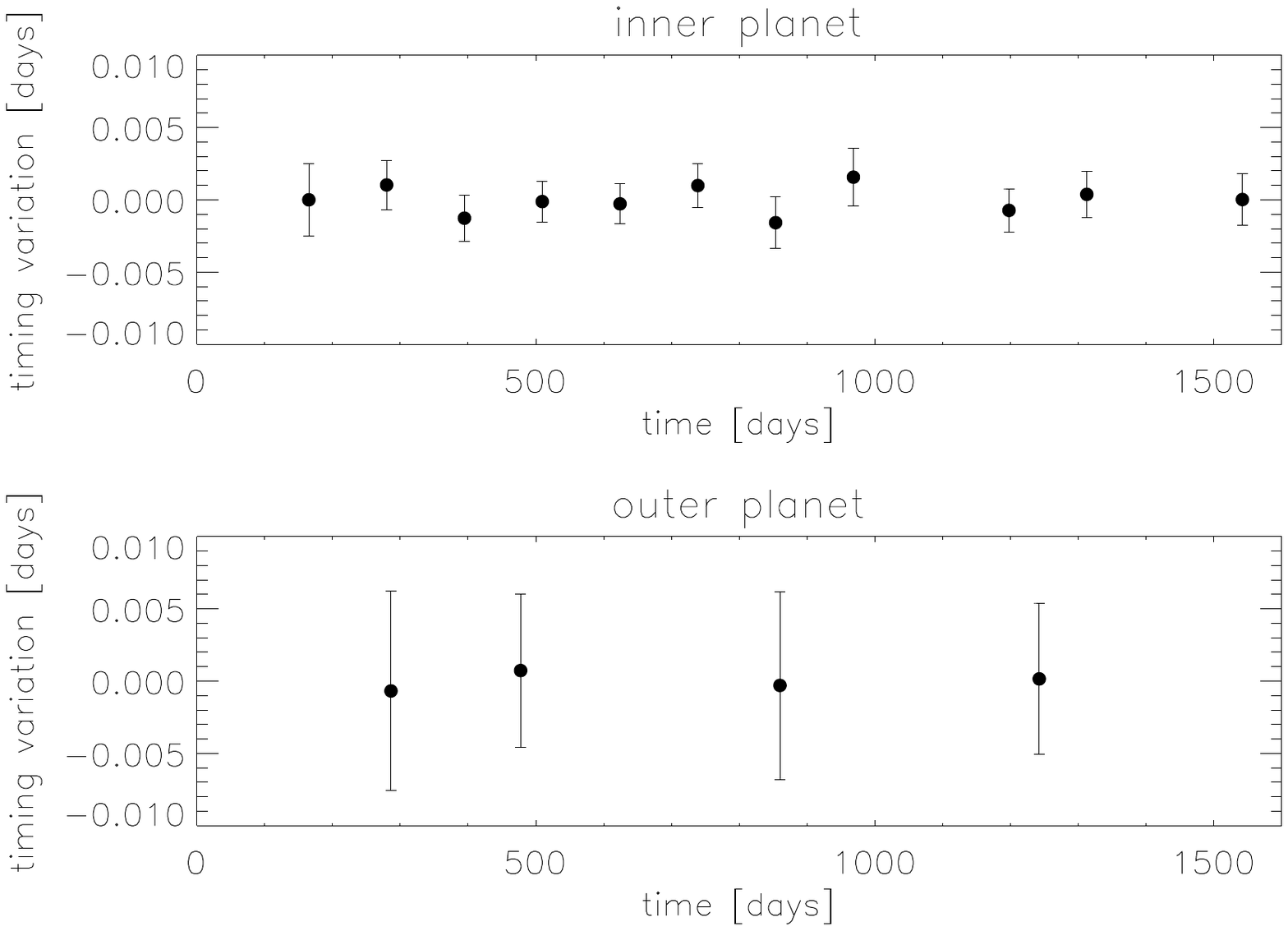}
\caption{Residuals between observed and calculated TTVs.}  
\label{Figttvres}
\end{figure}

The deviations of the observed and simulated transit timings of the fit are presented in Fig.\,\ref{Figttv} and Fig.\,\ref{Figttvres}, the latter showing the residuals. The reduced $\chi^2$ is 0.8.  We also risk and attempt to predict the times of mid-transit for the next few years (see Table \ref{Preditions}), but we note that the seemingly over-determined solution to Kepler-87\,c make it difficult for us to put reliable error bars on the prediction. We do that mainly since comparing future observations of such timings with the predicted ones given here offer the quickest way of checking if corrections to the model given in this paper are needed. This is quite likely, since the current observations cover the $\approx 550$\,days short term interaction cycle but do not cover the longer term interaction cycle of $> 3000$\,days. 

The minimum mutual Hill distance for the mean fit parameters is 4.7, which makes a long-term stable configuration plausible. Nevertheless, we integrated the orbit for 8 Gyr, i.e. the expected age of the system to ensure the dynamical stability, and found it to be stable.

\section{Discussion}
\label{Discuss}


We presented the dynamical confirmation of two long period low-density transiting planets using transit timing variation, as well as the initial detection of two transiting super-Earth sized planet candidates, all in the Kepler-87 system. Kepler-87\,c is the longest-period confirmed transiting planet around a single star, and Kepler-87\,b has the third-longest period, after the previous record holder Kepler-30\,d (Fabrycky \etal 2012). Batalha \etal (2013) strengthened the case for the paucity of short-period ($<10d$) giant planets in multiple systems (Latham \etal 2011). However, this paucity seems to be less severe for longer periods giant planets such as Kepler-87\,b. Particularly, the KOI 1574.04 planet candidate was not detected neither in the Q0-Q6 data (Batalha \etal 2013) nor in the Q1-Q12 data (Tenenbaum \etal 2012).

\begin{figure}[tbp]\includegraphics[trim=0 100 0 125, width=0.48\textwidth]{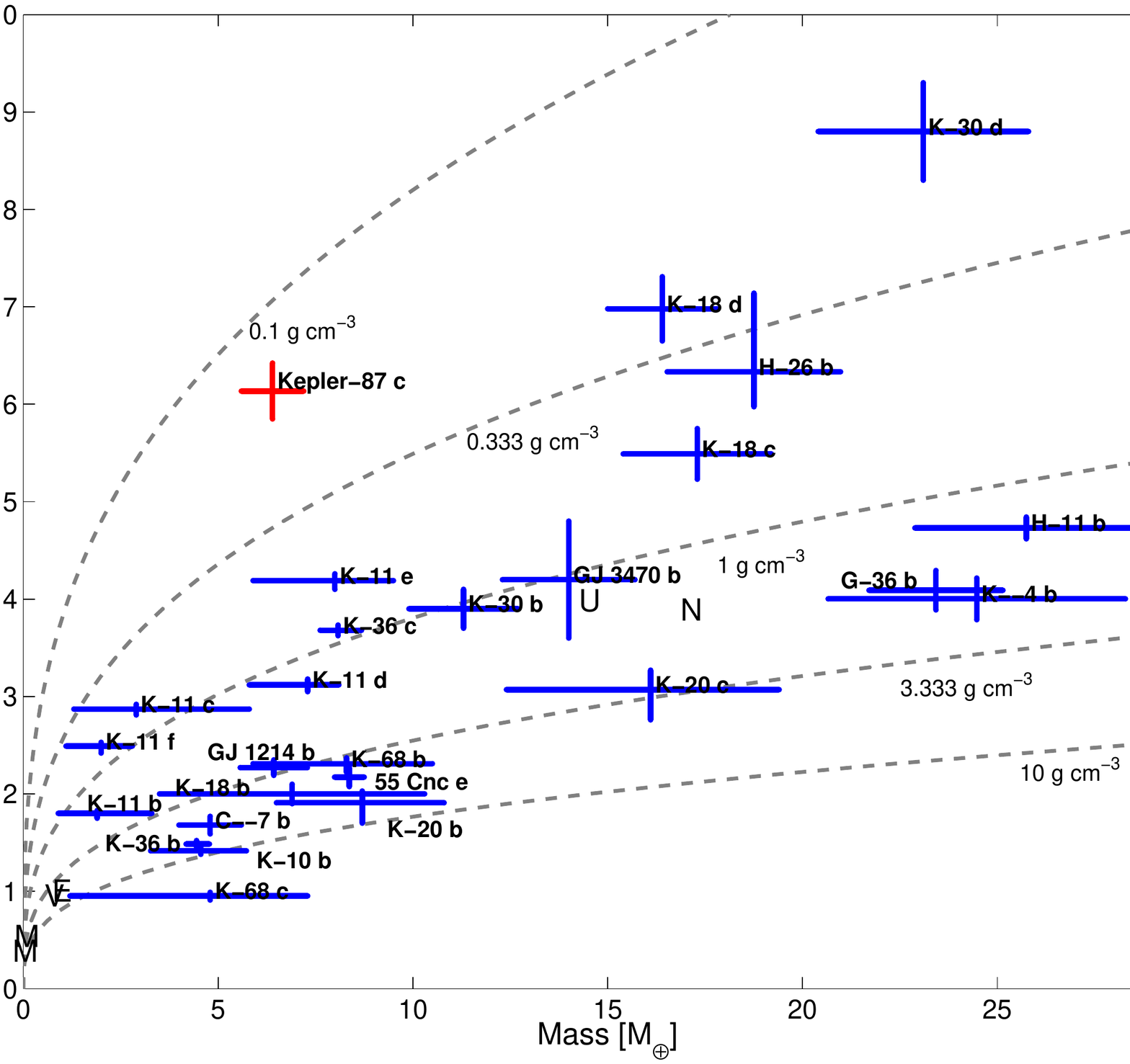}
\caption{Mass-radius relation for all known planets with masses below 30 $M_\oplus$ with overplotted bulk density contours. It is obvious that Kepler-87\,c occupies a unique position on this parameter space as the lowest-density planet for its super-Earth mass range. Some planet names were shortened so that ``K-X'' stands for the planet Kepler-X, and similarly ``C-X'' and ''H-X`` stand for ''CoRoT-X`` and ''HAT-P-X``. Solar system planets are designated with a letter with no error bars. We note Kepler-87\,b is beyond the scope of this figure (see discussion in the main text).}
\label{M_R}
\end{figure}

The most important feature of the Kepler-87 system is its two low density planets (Fig. \ref{M_R}). While Kepler-87\,b ($\rho_{01}=\RhoIn \pm \RhoInErr)$ has mass and radius that put is directly in the center of the general distribution of giant planets, Kepler-87\,c ($\rho_{02}=\RhoOut \pm \RhoOutErr)$ is anomalously low-density for its mass, similar to that of the least dense very hot Jupiters (e.g. Hartman \etal 2011). However, the Kepler-87 planets are rarefied despite the fact that they are neither strongly irradiated nor young. Such low densities suggest that a significant mass fraction can be attributed to Hydrogen and Helium. While common for giant planets, such a composition is non-trivial for planets less massive than $10 M_\oplus$ such as Kepler-87\,c which at no point in its history had the canonical critical mass for the starting of gas accretion of $10 M_\oplus$. Comparing these low densities to previously known planets is difficult since there are very few similarly long-period transiting planets, but the few known circumbinary planets already include Kepler-35\,b (Welsh \etal 2012) which has a density of $0.410^{+0.070}_{-0.069}\,g\,cm^{-3}$ - intermediate to the Kepler-87 planets. Importantly, there is no simple analogue to the low density of Kepler-87\,c.

Initially this result seemed baffling to us. However, the above solution exhibits strong self-consistency between different determinations of some of the variables: the stellar mass from spectroscopy agrees with the stellar mass from dynamical modeling, and the semi-major axes from light curve fitting + stellar model agree with semi-major axes from dynamical model. From a theoretical stand point, Rogers \etal (2011) attempted to put limits on the masses of similarly-sized \emph{Kepler} candidates and found that even low-mass low-density planets were possible in the general framework of core-nucleated accretion using plausible disc configurations. They found that a planet with a radius of $6 R_\oplus$ like Kepler-87\,c and an equilibrium temperature of $500 ^\circ\,K$ would have a mass of \MpOut $M_\oplus$ if $\approx20\%$ of its mass were made of a gaseous envelope (assuming ice-rock interior, and H/He in protosolar proportions). Correcting for the lower equilibrium temperature of Kepler-87\,c ($T_{eq}=$\TeqOut), its envelope mass fraction is probably somewhat higher than that. Planets such as Kepler-87\,c, as well as the highly irradiated Kepler -11 and -36 systems (Lissauer \etal 2011 and Carter \etal 2012), demonstrate that the great compositional variety that was found for gas giants also extends down to planets with masses intermediate between Earth and Uranus.

We believe that the two large planets of the Kepler-87 system present an opportunity for detailed study of exoplanet interior structure: residing at a relatively large orbital distance they are significantly less affected by the extreme insolation, that on shorter period planets may produce inflated radii on the one hand and mass loss due to irradiation driven atmospheric escape on the other hand. Furthermore, the host star is at a stage of its evolution that is relatively age-sensitive, making the system age relatively well determined. This relatively benign and constrained environment should make the two planets more amenable to modeling.

The end of the {\emph Kepler} mission also presents an interesting case were ground-based photometric follow-up of {\emph Kepler} planets is very desirable: the systematic uncertainty associated with the low number of data points means that additional observations are of significant value. On the one hand, Kepler-87\,b is an easy target (0.5\% depth) to better {\emph Kepler}'s 2-plus-minutes timing precision, and almost any-precision detection of Kepler-87\,c will be worthwhile due to its very large amplitude TTVs. On the other hand, due to their long periods and large host star both planets exhibit long transits of about 12hr and 15hr, which means full transits probably need a multi-site campaign. Note the accumulated effect of TTVs are very significant and can be even more than a day already in the near future (see Table \ref{Preditions}).

\section{Acknowledgements}

A.O. acknowledges financial support from the Deutsche Forschungsgemeinschaft under DFG GRK 1351/2. M.Z. acknowledges support by the European Research Council under the FP7 Starting Grant agreement number 279347. We thank Guillem Anglada-Escud\'e for discussing at length this system with us. We thank Bill Cochran and the rest of his team who observed KOI\,1574 and made the data available on CFOP. We would like to thank the team of the Hobby Eberly Telescope for taking the data of KOI\,1574. This paper includes data collected by the Kepler mission. Funding for the Kepler mission is provided by the NASA Science Mission directorate. Some/all of the data presented in this paper were obtained from the Mikulski Archive for Space Telescopes (MAST). STScI is operated by the Association of Universities for Research in Astronomy, Inc., under NASA contract NAS5-26555. Support for MAST for non-HST data is provided by the NASA Office of Space Science via grant NNX09AF08G and by other grants and contracts. This research has made use of the NASA Exoplanet Archive, which is operated by the California Institute of Technology, under contract with the National Aeronautics and Space Administration under the Exoplanet Exploration Program.
\bibliographystyle{plain}

\end{document}